\documentclass[aps,prl,superscriptaddress,reprint,amsfonts,amsmath]{revtex4-1}
\usepackage[final]{graphicx}
\usepackage{textcomp}

\let\oigr\includegraphics
\def\includegraphics[#1]#2{\IfFileExists{#2.eps}{\oigr[#1]{#2}}{\oigr[#1]{figures/#2}}}
\usepackage[usenames,dvipsnames]{xcolor}
\usepackage[normalem]{ulem}

\begin{document}
\title{Clearing out a maze: The hungry random walker and its anomalous diffusion}
\date{\today}
\def\dlr{\affiliation{Institut f\"ur Materialphysik im Weltraum,
  Deutsches Zentrum f\"ur Luft- und Raumfahrt (DLR), 51170 K\"oln,
  Germany}}
\def\hhu{\affiliation{Department of Physics,
  Heinrich-Heine Universit\"at D\"usseldorf,
  Universit\"atsstr.~1, 40225 D\"usseldorf, Germany}}
\def\unilu{\affiliation{Theory of Soft Matter, 
  Physics and Materials Science Research Unit,
  Universit\'e du Luxembourg, L-1511 Luxembourg, Luxembourg}}

\author{Tanja Schilling}\unilu
\author{Thomas Voigtmann}\dlr\hhu

\begin{abstract}
We study chemotaxis in a porous medium using as a model a biased (``hungry'')
random walk on a percolating cluster. In close resemblance to the
1980s arcade game Pac-Man\textsuperscript{\textregistered}, the
hungry random walker consumes food, which is initially distributed in the maze,
and biases its movement towards food-filled sites.
We observe that, on the percolating cluster, the
mean-squared displacement of the pacman process shows anomalous dynamics, which
follow a power law with a dynamical exponent different
from both that of a self
avoiding random walk as well as that of an unbiased random walk.
The change in dynamics with the propensity to move towards food
is well described by a dynamical exponent that depends continuously
on this propensity, and results in slower differential growth when compared
to the unbiased random walk.
\end{abstract}

\maketitle

\begin{figure}
\includegraphics[width=.9\linewidth]{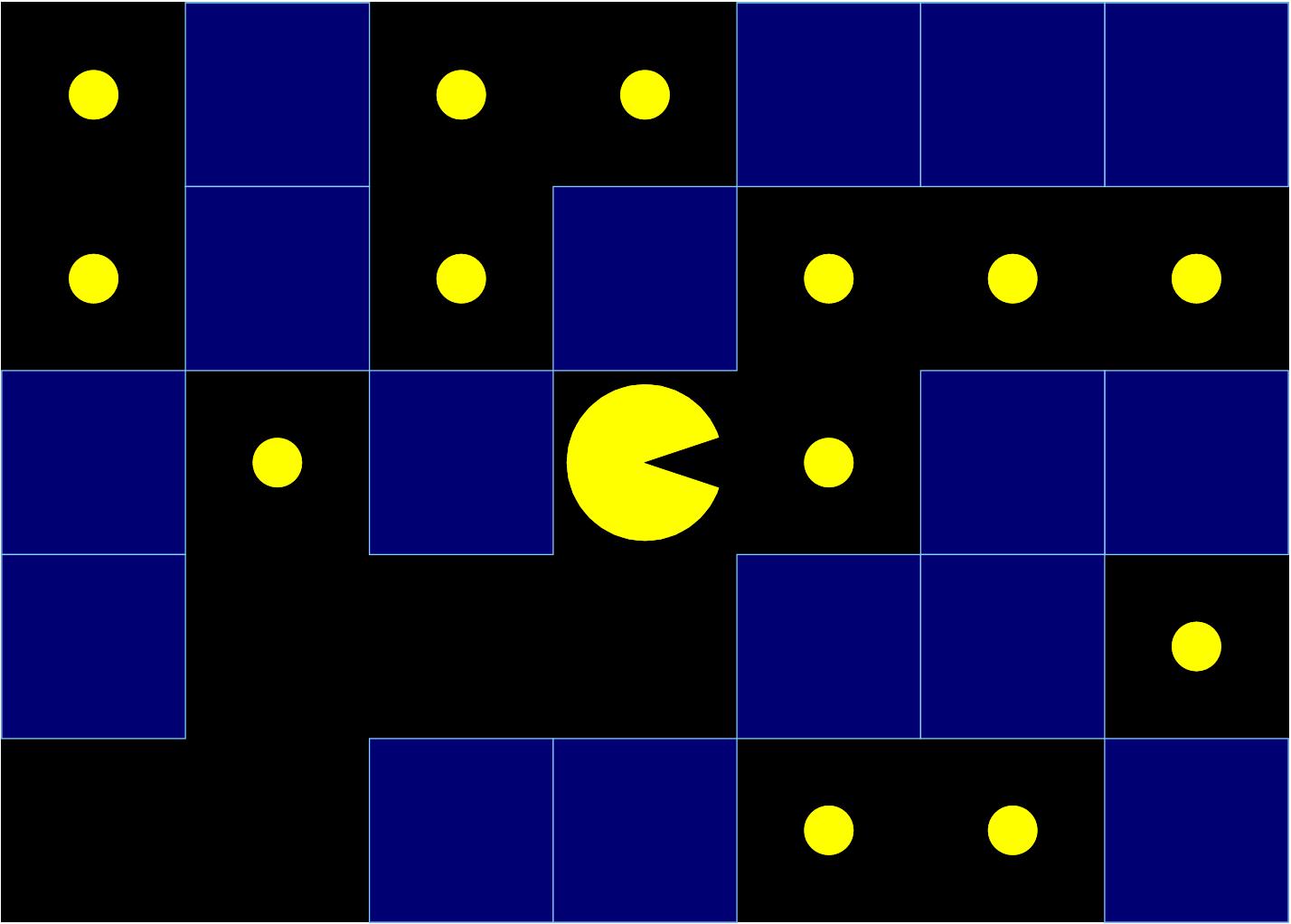}
\caption{\label{fig:setup}
Sketch of the model. Blue squares are obstacles, small yellow circles are food and the large yellow symbol is the random walker (pacman).
}
\end{figure}

Consider a random walker confined by a disordered environment of obstacles 
(see Fig.~\ref{fig:setup}).
Let the walker be free to move in the void space between the obstacles, with its movement biased to nearby places where it encounters a resource (``food''), which it then consumes.
On a two-dimensional square lattice the sites of which are randomly blocked (with probability $1-p$), this process resembles the motion of Pac-Man\textsuperscript{\textregistered} known from the famous 1980's arcade game. 

The unbiased random walk on such a lattice with obstacles is one of the simplest models for transport in porous media \cite{Stauffer.1991,Havlin.1987}. It is known to show ``anomalous'' dynamics, i.e.~non-trivial power laws in the transport properties that are caused by the fractal structure of the accessible sites. When the probability for a site not to be blocked, $p$, crosses the critical probability $p_c$, a system-spanning (``percolating'') cluster of accessible sites first appears in the infinite system. The mean-squared displacement (MSD) of an unbiased random walker that is confined to the percolating cluster follows the asymptote $\delta r^2(t)\sim t^{2\nu}$ with an exponent $\nu<1/2$ given by the walk dimension $d_w$, $\nu=1/d_w$. For the 2D square lattice, $d_w\approx2.878$ and $p_c\approx0.592746$ \cite{Grassberger.1999}. (In the following we will reserve the term ``random walk'' to refer to this unbiased walk.)

Following a food bias resembles the chemotactic motion of bacteria in porous media. The dynamics of microswimmers in crowded environments of obstacles is of obvious importance for the physics of biological systems \cite{Chepizhko.2013,Chepizhko.2013b}, and also for applications such as groundwater decontamination (where food to the bacteria is an unwanted pollutant) or microbial enhanced oil recovery (where microbes feed on long hydrocarbon chains, breaking them up into valuable light crude oil) \cite{Patel.2015}.
Two aspects of bacterial motion are important in this context: the first is the bacteria's starvation behavior, either because they die, or because starved bacteria tend to increase their efficiency in seeking food. The second is the interplay between anomalous diffusion and chemotaxis, in particular concerning the exploration (and subsequent clogging) of dead ends in the percolating structure.
Lattice-based random-walk models of starving walkers have recently been discussed \cite{Benichou.2014,Chupeau.2016}, albeit in the homogeneous medium. Our model focuses on the second aspect, the chemotactic exploration of the porous medium. 

We model the motion as a discrete process on the lattice. At every time step, the walker at position $i$ has a probability to move to any of its allowed nearest-neighbor sites $j$ given by
\begin{equation}\label{eq:prob}
  p_{j\leftarrow i}=\frac{\exp F_j}{\sum_{k\sim i}\exp F_k}\,,
\end{equation}
where $F_k$ is the propensity to move towards food at site $k$, $F_k=0$ if $k$ is blocked,
and the sum runs over all nearest neighbor sites \footnote{If $j$ is a site blocked by an obstacle $p_{j\leftarrow i}=0$, resulting in a finite probability to remain at site $i$, as in de~Gennes' ``blind ant'' model \protect\cite{Havlin.1987}.}. Initially, all accessible lattice sites are assigned a homogeneous propensity $F$. As the process evolves, food is removed from all the sites $j$ that the walker encounters, setting $F_j=0$. This describes a non-Markovian transient process that clears out an increasingly large portion of the accessible space.
In this sense, it is a near-sighted version of Pac-Man\textsuperscript{\textregistered}. (We do not attempt to model the behaviour of a human player who would bias his steps based on the known structure of the maze \footnote{For simplicity, we also do not consider ghost modes.}.)

At first glance, the model is quite similar to generalizations of the self-avoiding walk (SAW). In the SAW the walker is free to move to any not-yet visited site, but never allowed to revisit an already encountered site. On a lattice without obstacles (free SAW), its MSD grows as $\delta r^2(t)\sim t^{2\nu}$ with an anomalous exponent $\nu^\text{SAW}>1/2$, i.e.~faster than the free random walk. In 2D, $\nu^\text{SAW}=3/4$ has been conjectured as the exact value \cite{MadrasSlade, Lawler2004}. The SAW on the percolating cluster is a model for linear polymers in heterogeneous media. Its exponent $\nu^\text{pcSAW}$ is larger than that of the free SAW \cite{Fricke.2014}, specifially $\nu^\text{pcSAW}\approx0.78$ in 2D \cite{Blavatska.2008,Fricke.2012}. Thus, while the random walk slows down when constrained to the percolating cluster, the SAW speeds up.

Variants of the SAW that merely suppress repeated visits to the same site instead of disallowing them are quite similar to our model \eqref{eq:prob}. They have been termed ``true'' \cite{Amit.1983} or myopic self-avoiding walks \cite{Lawler.1991}. On the percolating cluster, their MSD are also found to grow faster than that of the random walk \cite{Bouchaud.1989,Lee.1990,Lee.1992}.

The surprising result from our simulations is that the MSD of the hungry walker on the percolating cluster grows \emph{less} efficiently than that of the random walker at long times: it is described by a different walk dimension $d^\text{pacman}_w$ that is larger than that of the unbiased random walk, $d_w$ (i.e.~by an exponent $\nu^\text{pacman} < 1/2.878$).

We have performed Monte Carlo simulations of the process described by Eq.~\eqref{eq:prob}. Random matrices for $p=0.592746$ were created with lateral dimensions between $L=10000$ and $L=25000$, and periodic boundary conditions were employed for the trajectories.
Wrapping site-percolating clusters were identified by the Hoshen-Kopelman algorithm \cite{Hoshen.1976}. The random walks were started on a random location on the (largest) percolating cluster, and results are typically averaged over $90000$ runs of up to $5\times10^8$ steps each (using $N=100$ walks with random starting points on each of $M=900$ matrix realizations).
Jump probabilities are calculated by keeping track of already visited sites in a hash table \cite{Madras.1988}.
We have checked for finite-matrix-size effects in some runs up to $10^{10}$ steps \footnote{On average, the process for $F=10$ consumes approximately $3333360$ units of food in $10^{10}$ steps, coincidentally the maximum score achievable in the original Pac-Man arcade game, cf.\ \protect{http://heise.de/-3227685}.} and in matrices with sizes up to $L=120000$. The robustness of our long-time extrapolations was checked to be compatible with a few runs up to $10^{12}$ steps for $F=10$.

\begin{figure}
\includegraphics[width=.9\linewidth]{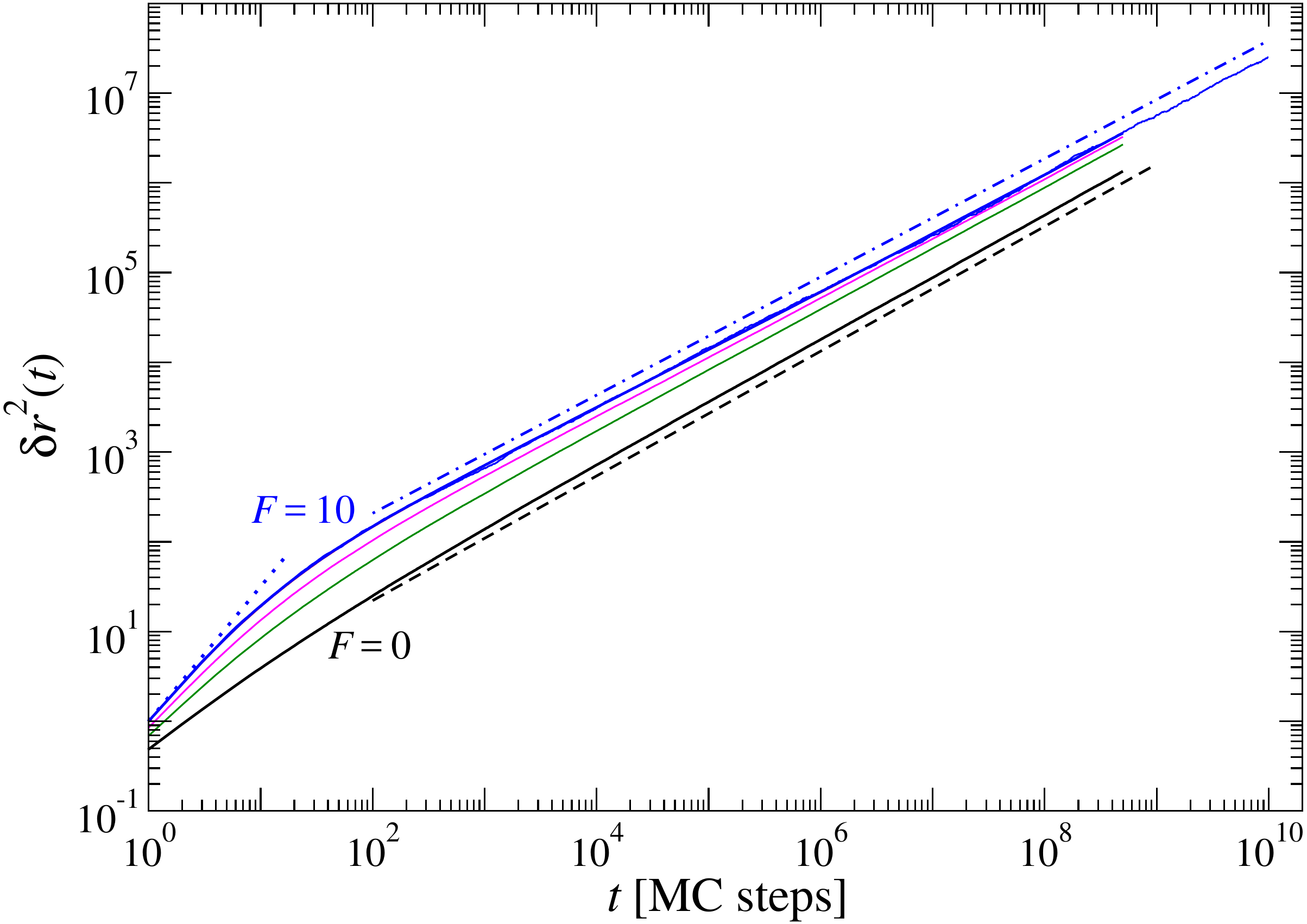}
\caption{\label{fig:msd}
Mean-squared displacement (MSD) $\delta r^2(t)$ of the hungry random walker on the 2D site-percolating cluster for food propensities $F=0$, $1$, $2$, $5$ and $10$ (the last two coinciding, from bottom to top). For $F=10$, an average over $1000$ runs up to $t=10^{10}$ is also shown. The dashed line is the asymptotic power law for $F=0$, $\sim t^{2/d_w}$ with $d_w=2.878$; the dash-dotted line is $\sim t^{2/d_w^\text{pacman}}$ with $d_w^\text{pacman}=3.03$. The dotted line indicates $t^{3/2}$ at small $t$.}
\end{figure}

Figure~\ref{fig:msd} shows the MSD obtained for walks on the percolating cluster with different food propensities. The case $F=0$ (random walk) is included for reference. It displays the expected asymptotic law $\delta r^2\sim t^{2/d_w}$. The MSD with $F>0$ increase faster initially: After the first step, the walker remains strongly biased towards exploring new sites. Indeed, for large $F$, the initial increase in the MSD is $\sim t^{3/2}$, the power law expected from the free SAW, since at short times, the fractal structure of the percolating cluster is not yet explored. After this transient,
anomalous subdiffusion manifests itself.
Remarkably, the power-law increase of the $F>0$ curves is
\emph{slower} than that of the $F=0$ reference. Up to the longest time accessible
in our simulations, $t=10^{10}$, the MSD for $F=10$ appears to be well
described by a power law $\delta r^2\sim t^{2/d_w^\text{pacman}}$ with an exponent $d_w^\text{pacman}$
that is larger than $d_w$. This is emphasized by the dash-dotted line in
Fig.~\ref{fig:msd} that represents the power-law with $d_w^\text{pacman}=3.03$,
the exponent that best describes our $F=10$ data.

\begin{figure}
\includegraphics[width=.9\linewidth]{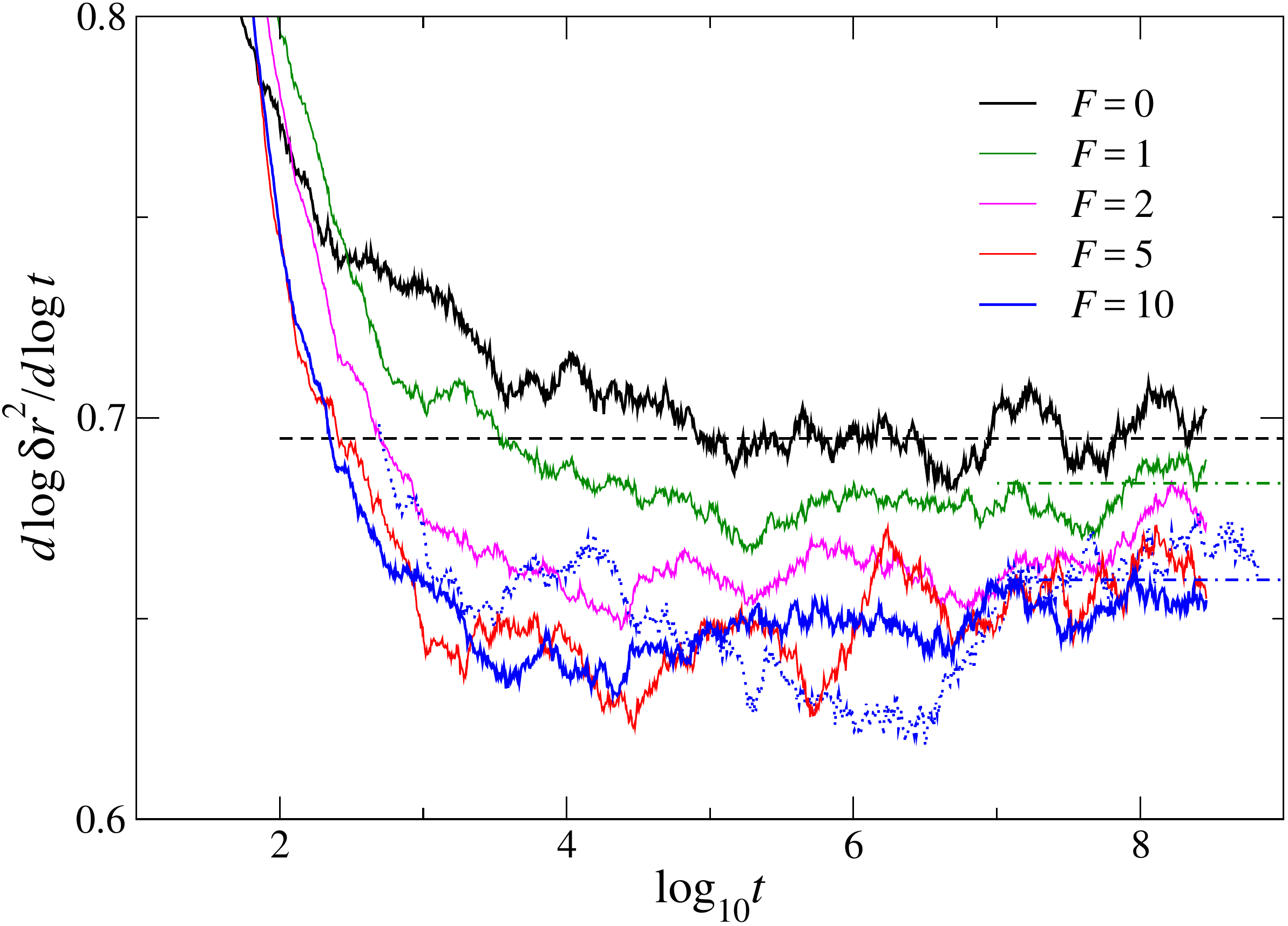}
\caption{\label{fig:dlogmsd}
Effective exponent $\alpha(t)=d\ln\delta r^2/d\ln t$ corresponding to the
MSD shown in Fig.~\ref{fig:msd}. The horizontal dashed line indicates $2/d_w$, dash-dotted lines indicate average values of $\alpha(t)$ for $t>10^7$. The data was smoothened by a moving average after logarithmic differentiation.
}
\end{figure}

The power-law growth becomes clearer when looking at the logarithmic derivative of the
mean-squared displacement, $\alpha(t)=d\ln\delta r^2/d\ln t$. If the MSD
follows a power law, $\alpha(t)$ is a constant and its value is the power-law
exponent. As shown in Fig.~\ref{fig:dlogmsd}, the effective exponent $\alpha(t)$
for $F=10$ indeed
remains close to $2/d_w^\text{pacman}$, and stays significantly below the value $2/d_w$ expected for $F=0$, over 6 orders of magnitude in time.
Remarkably, the thus estimated exponent $d_w^\text{pacman}$
is close to the known value of the dynamical
critical exponent for the random walk, $z\approx3.036$,
obtained from the all-cluster averaged MSD \cite{Grassberger.1999} (i.e.~for a set of walkers that starts out on any non-blocked site rather than just the sites of the percolating cluster). 
Recall that for the latter, trajectories contribute
that eventually localize, with a weight given by the cluster-size distribution
\cite{Havlin.1987,Kammerer.2008};
this leads to $z=2d_w/d_f$ in 2D, where $d_f=91/48$
is the fractal dimension of the percolating cluster.

Our finding can be rationalized by the following conjecture:
any random walk on the percolating cluster will perform random excursions
that leave the backbone of the cluster and explore culs-de-sac out of which
the process has to return. These culs-de-sac are dangling clusters that
have a fractal size distribution (the same as the overall cluster size
distribution \cite{Porto1999}). While the random walk on the 
percolating cluster sees some effect of these dangling clusters, the 
effect of food bias is
to draw the walker more deeply into any cul-de-sac whose entrance it randomly
samples. Thus, over sufficiently long time scales, the ``pacman'' process
will perform excursions that slow down the overall growth of the MSD as 
compared to the random walk.

That the dangling ends have an effect on the walk dimension is known from the case of the random walk in 2D. If the walker is restricted to explore only the backbone of the percolating cluster, $d_w^\text{bb}\approx2.70<d_w$ \cite{Blavatska.2009}, i.e., the MSD grows faster than on the full cluster.
\footnote{For the SAW on the contrary, recent enumeration studies in 
3D suggest that the dangling ends are not relevant for the asymptotic 
growth of the MSD\cite{Fricke.2014}.}

\begin{figure}
\includegraphics[width=.9\linewidth]{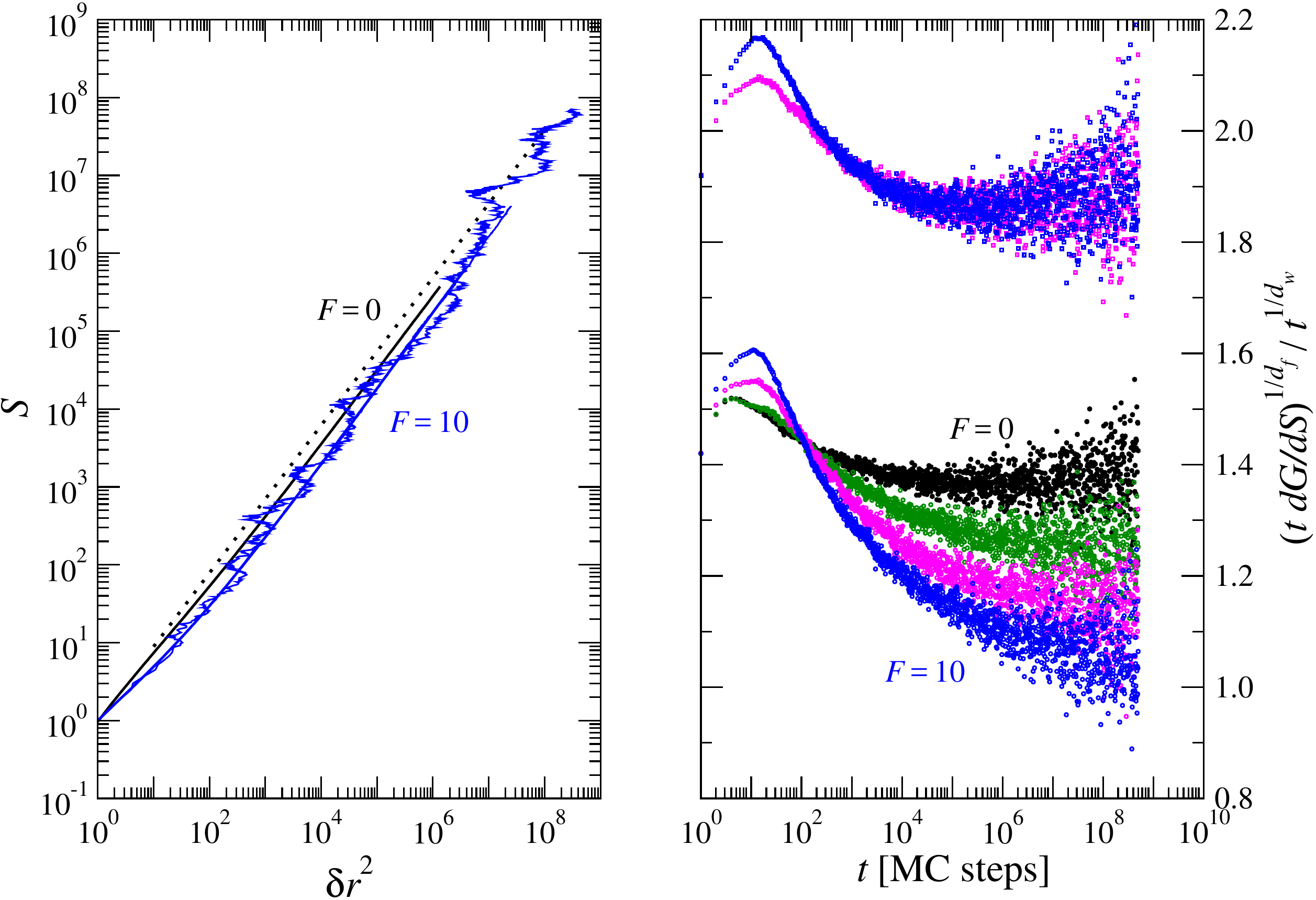}
\caption{\label{fig:st}
  (left) Number of distinct sites visited, $S$, as a
  function of the mean-squared displacement $\delta r^2$,
  for $F=0$ and $F=10$ (as labeled). The jagged curve for $F=10$ is
  a rough estimate for the $F=10$ case from $6$ individual runs up to
  $t=10^{12}$. The dotted line indicates $(\delta r^2)^{d_f/2}$.
  (right) Rectification plot for the growth rate of the process towards
  new sites, $dG/dS$, for $F=0$, $2$, $5$, and $10$, assuming $d_w=2.878$
  (lower set of curves, top to bottom), and assuming $d_w(F)$ to depend
  on $F$ (upper set of curves, $F=2$ and $F=10$).
}
\end{figure}

To further test whether the pacman dynamics is indeed described by a power law
with an exponent different from $d_w$, let us discuss the number of distinct
sites $S$ visited by the walker as a function of time.
For a random walk that is homogeneous, ergodic, and fulfills detailed
balance, one expects this number to scale with the dimensionality of the
available space, i.e., $S\sim R^{d_f}$ for a walk of typical lateral
extension $R$.
The left panel of Fig.~\ref{fig:st} shows $S$ as a function of $\delta r^2$
for the extreme cases $F=0$ and $F=10$.
For $F=0$, the curve exhibits the expected scaling law
$S\sim(\delta r^2)^{d_f/2}$ (indicated by a dotted line),
where $(\delta r^2)^{1/2}\sim R$ is a good proxy of the size of the
spherical area covered by the walk.
The process for $F=10$ visits more distinct sites than the 
random walk in the amout of same time, but to reach the same number 
of sites $S$ it spreads out farther than the random walk. 
Hence, its $S$-versus-$\delta r^2$ curve
stays below that of the random walk. Interestingly, a closer analysis of
the data shown in Fig.~\ref{fig:st} reveals that for times $t>10^4$ and
up to at least $t=10^{10}$,
the quantity $S$ grows faster than $(\delta r^2)^{d_f/2}$. Since
$S\sim R^{d_f}$ is an upper bound for the growth law, this implies that
(at least in this time window) the MSD is not a good indicator of the
typical area covered by the walk. A possible explanation in line with our
rationale of the slower growth observed in the MSD is that indeed,
the pacman process first traverses the dangling ends of the percolating
cluster in a low-dimensional fashion similar to a SAW, followed by a
more space-filling exploration of the dangling ends that leads to the
faster growth in $S$ with $\delta r^2$ that we observe. Since dangling
ends have a power-law size distribution, a power-law-like increase in $S$
results.

Still, the observed behavior in $S(\delta r^2)$ might suggest a cross-over
to ordinary random-walk dynamics at times much larger than those that we can
access in our simulation, and that the values of $\alpha(t)$ discussed in
Fig.~\ref{fig:dlogmsd} are the result of such a cross-over.
We therefore present a further estimate of $d_w^\text{pacman}$ based on
the growth of $S$. Following an argument by Leyvraz and Stanley
\cite{Leyvraz.1983}, consider the average number $G$ of possible growth sites
for $S$ at a given time, i.e., the average number of unvisited nearest-neighbor
sites. Since $G$ only changes when $S$ changes, one would estimate
$dG/dS\sim S^{-1/2}$ if $G$ were composed of identically independently
distributed random increments. Assume that relevant correlations (introduced both by the
structure of the percolating cluster and by the non-Markovain jump rates)
change this to $dG/dS\sim S^{-x}$ with some $x\neq1/2$. The rate of
increase in the number of sites visited $S$ will scale with the probability
to step onto a growth site; one thus estimates $S/(dG/dS)\sim t$, which
is well fulfilled by our numerical data.
Assuming now that asymptotically, $S\sim t^{d_f/d_w'}$, plus
corrections to scaling that are likely to cancel out in $S/(dG/dS)$,
we get $\mu(t)=t\,dG/dS(t)\sim t^{d_f/d_w'}$, where $d_w'=d_f(1+x)$ is the
walk dimension of the process to be determined.
We estimated
$dG/dS$ by counting the average number of unvisited nearest-neighbor sites
seen at every step, and show $\beta(t)=\mu(t)^{1/d_f}/t^{1/d_w'}$ in the
right panel of Fig.~\ref{fig:st}. The lower set of curves assumes $d_w=2.878$
and exhibits the expected constant that signals the scaling regime only
for $F=0$ for times $t\gtrsim10^4$. For all $F>0$, no such asymptote
is visible. In particular, there is, in the time window $t<5\times 10^8$,
no obvious $F$-dependent time scale
that would indicate the cross-over from an intermediate-time regime to
the random-walk asymptote.
Taking for $d_w'$ those values found from the logarithmic derivative of the
MSD, Fig.~\ref{fig:dlogmsd}, we obtain also for $F>0$ constant values for
$\beta(t)$ at large times. This is demonstrated by the upper set of curves
in the right panel of Fig.~\ref{fig:st}. Note that the average $dG/dS$ is
dominated by rare events for large $S$ and thus the results for $\beta(t)$
have large error bars at late times. This also indicates that at those late
times, deviations from the proper asymptote in our data may be due to
insufficient sampling of trajectories. This can explain the slight
upward bend of the curves shown in Fig.~\ref{fig:dlogmsd} because the
transition rates in Eq.~\eqref{eq:prob} might be numerically biased towards
those of the ordinary random walk.

\begin{figure}
\includegraphics[width=.9\linewidth]{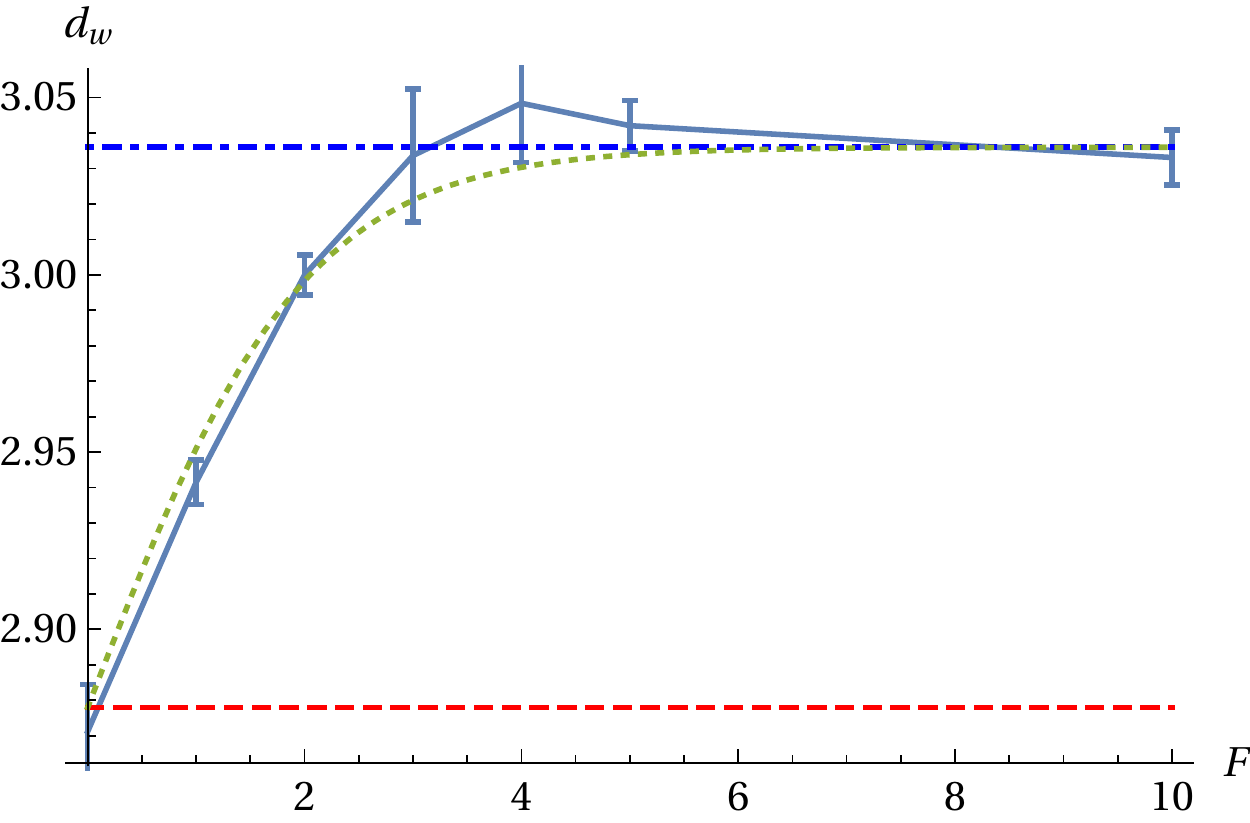}
\caption{\label{fig:fexp}
  Power-law exponent $\alpha$, shown as the walk dimension $d_w(F)=2/\alpha$
  for the MSD of pacman obtained from averaging
  $d\ln\delta r^2/d\ln t$ for times $t\gtrsim1.4\times10^7$, as a function
  of food propensity $F$. Horizontal lines indicate the known values
  for $d_w\approx2.878$ and $z\approx3.036$. Error bars indicate standard
  deviations from different realizations for the fixed time interval.
}
\end{figure}

While we can only observe a finite time window in the simulation, it is
conceivable that at large $F$, the long-time asymptote of the MSD is
indeed different from the one at $F=0$. Since the number of unvisited
nearest-neighbor sites seen by the ensemble of walkers at time $t$ decreases
slowly as a power law, the pacman process remains intrinsically
non-stationary: there appears to be no time scale where the walker has
exhausted all food in its vicinity, and thus the number of biased steps
never vanishes.
In this respect, our model is very different from that adopted for
``starving walkers'' \cite{Benichou.2014}:
even when encountering food-filled nearest neighbors,
the starving walker remains unbiased, so that long-lived starving walkers are
expected to follow the power law $t^{2/d_w}$ obtained for the random walk
on the percolating cluster.

Since there is no obvious intrinsic time scale characterizing the $F$-dependence
of the motion, the observed exponent itself can change continuously with $F$.
In Fig.~\ref{fig:fexp} we show the values of $\alpha(F)$ estimated from our
simulations, obtained by averaging the logarithmic derivatives of the MSD
for times $t\gtrsim1.4\times10^7$. The values
depend somewhat on the time window chosen for averaging, but the continuous
increase from $F=0$ to $F\to\infty$ is robust.
Such a continuously changing exponent is best rationalized by observing
that the efficiency with which the walker is drawn into the culs-de-sac
also depends on $F$. As $F$ increases, the size-distribution of dangling
ends is probed with a weight function that depends on $F$. 
In this respect, our model differs from, e.g., the change in universality
classes incurred by a change in microscopic dynamics in the three-dimensional
Lorentz model \cite{Spanner.2016}.

Our statement that this kind of model shows a continuously changing 
exponent, is also supported by the case of the hungry walker on the 
one-dimensional lattice with an absorbing boundary 
at the origin, for which the survival probability has been shown to scale 
as a power law in time, with an exponent that depends on $F$ \cite{Dickman.2001}.

In summary, we have studied a model for chemotaxis in a random medium that resembles the game Pac-Man\textsuperscript{\textregistered}: the non-Markovian random
process of moving a food-consuming  (hungry) walker on the percolating 
cluster on a 2D square lattice. It shows
anomalous sub-diffusion that is well described by a propensity-dependent
dynamical exponent. Its mean squared displacement follows 
a  power-law that is \emph{slower} than that of the unbiased walk.
 For increasing $F$, the
walker initially moves farther from the origin than the random walk, but
then becomes less effective in exploring area. We argue that this
results from the fact that the hungry walker ``gets lost'' in
the culs-de-sac: it tends to explore
the dangling ends of the percolating cluster in more depth than the unbiased
walker. Qualitatively, this matches the observation that chemotactic bacteria
tend to get stuck in the ``dead zones'' of porous media.
The dependence of the dynamics on food propensity is better described
by assuming a $F$-dependent
exponent, rather than by a cross-over behavior between two fixed power laws.
Our model hence provides a simple example of continuously changing
dynamical exponents.

\begin{acknowledgments}
We thank T.~Franosch and G.~Peccati.
\end{acknowledgments}

%

\end{document}